    \documentclass[prl,aps,twocolumn,showpacs]{revtex4}
\usepackage[dvips]{graphicx}
\usepackage{dcolumn}%
\usepackage{amsmath}%
\usepackage{amsfonts}%
\usepackage{amssymb}
\providecommand{\U}[1]{\protect\rule{.1in}{.1in}}
\newcommand{\be}{\begin{equation}}
\newcommand{\ee}{\end{equation}}
\newcommand{\ba}{\begin{array}}
\newcommand{\ea}{\end{array}}
\newcommand{\nn}{ \nonumber}
\newcommand{\ds}{\displaystyle}
\begin{document}

\title{Electric charge and potential distribution in twisted multilayer graphene}

\author{Natalya A. Zimbovskaya$^{1,2}$ and Eugene J. Mele$^3$}

\affiliation
{$^1$Department of Physics and Electronics, University of Puerto 
Rico-Humacao, CUH Station, Humacao, Puerto Rico 00791, USA; \\
$^2$Institute for Functional Nanomaterials, University of Puerto Rico, San Juan, Puerto Rico 00931, USA}
\affiliation
{$^3$Department of Physics and Astronomy, University of Pennsylvania, Philadelphia, Pennsylvania 19104, USA}

\begin{abstract}
The specifics of charge screening and electrostatic potential spatial distribution in  multilayered graphene films  placed in between charged substrates is theoretically analyzed.  It is shown that by varying the areal charge densities on the substrates  and/or the thickness of the graphene stack one may tune the doped carriers distribution over the system. When the charge densities on the substrates are weak, the carriers distribution and electrostatic potential profile agree with semimetallic properties of graphene. However, when the amount of the donated charge is sufficiently large  the transition to a metallic-like behavior of the graphene layers occurs. 
 The possibilities for experimental observation of the predicted transition are discussed.
   \end{abstract}

%%{71.18.+y, 71.20-b, 72.55+s}

\date{\today}
\maketitle

\section{I. Introduction}

Theoretical and experimental studies of graphene last more than a decade, and the interest of the research community to these studies remains inabated. This interest is steming from remarkable electronic properties of graphene and graphene derived materials. In a single graphene sheet, the electronic properties are controlled by two-dimensional single-particle spectra containing linearly dispersed bands. 
    The inherent two-dimensionality of charge-carriers spectra in graphene suggests that it may have high potentialities for nanoelectronic applications. Multilayered graphene  films   are considered as potentially suitable materials for building nanoelectronic devices with planar geometry \cite{1,2,3}. 
  A MLG film supported by an insulating substrate or sandwiched in between two substrate layers should be an important element and building block in manufacturing of these devices. The substrates may affect electronic properties of MLGs placed in close proximity to them, especially when they are bearing electric charges thus creating  electric fields across the film.

 Correspondingly, the effects of the substrate on the electronic properties of MLGs must be thoroughly analyzed. Some theoretical and experimental works concerning this issue already exist (see e.g. Refs. \cite{4,5,6,7,8}). However, further studies are necessary to get better quantitative understanding of these effects. 
   In the present work we contribute to these studies by theoretically analyzing the charge exchange and electrostatic potential spatial distribution in the system, which consists of a   MLG film sandwiched in between  two charged substrate layers. 
     We show that the profile of the electrostatic potential strongly depends of the magnitude of electric charge on the substrates. When these charges are sufficiently small, the MLG behaves as an insulating/semimetallic material with a typical linear profile of the electrostatic potential and the screening length of the same order as the MLG film thickness. At higher values of the electric charge on the substrates, the potential profile changes acquiring features typical for metallic-like materials, and the electric charge induced in the MLG becomes accumulated near the interfaces separating the latter from the substrates. We analyze necessary conditions for this transition to occur and discuss possibilities for observation of this effect in experiments.

\section{II. Main Equations}

We consider a  MLG film
which occupies the space where $ -D/2 < z < D/2. $ The number of layers is supposed to be large enough to satisfy the condition $ d \ll D $ where $ d $ is the distance between adjacent layers in the film. At the interfaces between the MLG and the substrates $(z =-D/2,D/2, $ respectively) the latter are characterized with the areal charge carriers densities $ \sigma_{1s} $ at $ z = -D/2 $ and $ \sigma_{2s} $ at $ z = D/2. $ In the presence of charged substrates, the additional charge carriers appear at the graphene sheets as a result of charge transfer process. We introduce areal densities of these doped charges $ \sigma_i $ corresponding to the graphene layers in  the MLG. The index $ "i" $ takes on values from $ 1 $ to $ N \ (N $ being the total number of the layers in the pack). For convenience, in further analysis we  assume that all above mentioned charge carrier densities may take on either positive or negative values depending on the nature of the charge carriers associated with a certain layer/interface. We attribute positive values to the areal densities of holes and negative ones for those of electrons, respectively.
   We assume that the local Fermi energy characterizing a single sheet is sufficiently large, so we may use a conical dispersion relation for the doped charge carriers. By using this relation, we imply that the carriers density of states remains basically unchanged notwithstanding the electric field  created by the substrates. 
Due to the specific form of the dispersion relation for the doped charge carriers in a single graphene sheet, their quantum-mechanical kinetic energy per unit area is proportional to the areal charge-carriers density in power $ 3/2:$ 
\be
K = \frac{2}{3}\sqrt \pi \hbar v_F |\sigma|^{3/2}   \label{1}
\ee
where $ v_F $ is the Fermi velocity of the charge carriers. We remark that this expression differs from the well known result applicable to a conventional two-dimensional conducting system where $ K \sim \sigma^2. $
  Similar expression could be written for the kinetic energy associated with the $i $-th graphene sheet included in the pack assuming that there is no coupling between the adjacent sheets. However,
   electronic properties of multilayered graphene films (MLG) are more complicated than those of a single graphene sheet. When several sheets are put together, the coherent interlayer tunneling of charge carriers occurs. The effects of interlayer tunnelings could be quite strong provided that crystalline lattices of adjacent layers are arranged in some special way with respect to each other \cite{9,10,11,12}. To a significant degree, the interlayer motions may destroy the  two-dimensionality of the original graphene sheets.
    Moreover, the coupling between the layers in MLG films may lead to radical changes  of charge-carriers low energy spectra. Depending on separation distances between the adjacent layers and mutual orientations of their crystalline lattices (twist angles), the very topology of low-energy isoenergetic surfaces associated with the layers could undergo changes along with the Fermi velocity and other important transport characteristics. Presently, low-energy characteristics of rotationally faulted MLGs are intensely studied. A review of recent theoretical results concerning this issue is presented in Ref. \cite{13}.

Complicated form of low-energy dispersion relations for the charge carriers associated with a graphene sheet belonging to a MLG pack distinguishes them from simple dispersion relations corresponding to Dirac cones. This puts in question the validity of the expression for the kinetic  term (\ref{1}) when one considers a graphene sheet stacked together with other sheets to compile a MLG system. To estimate the scope of suitability of the expression  (\ref{1}), we turn to the case (repeatedly discussed  in existing papers) of a graphene bilayer with its two layers arranged in such a way that ``A'' sublattice of the top layer is placed precisely on the top of ``B'' sublattice of the bottom one (so called Bernal stacking geometry). This geometry is one of the most favorable for manifestations of interlayer coupling to occur, which determines our choice. It was shown that within the considered geometry four nondegenerate bands corresponding to the bilayer emerge \cite{14,15}:
\be
E_{\bf k} = \pm \sqrt{\epsilon_{\bf k} + 2t^2 \pm \sqrt{\epsilon_{\bf k}^2 t^2 + t^4}} . \label{2}
\ee
Here, $ \epsilon_{\bf k} $ are eigenvalues of the intralayer Hamiltonian in the absence of interlayer interactions, and  the parameter $ t $ has the dimensions of energy and characterizes the interlayer coupling strength. Solving Eq. (\ref{2}) for $ \epsilon_{\bf k}^2, $ one may  find expression for the charge-carriers densities of state corresponding to the nondegenerate bands and then compute their contributions to the kinetic energy term for a single layer. It may be approximated as follows:
\be
K \approx \frac{2}{3}\sqrt\pi \hbar v_F \left\{\left(|\sigma| + \frac{4t^2}{\pi\hbar^2v_F^2} \right)^{3/2} - \left(\frac{4t^2}{\pi\hbar^2v_F^2}\right)^{3/2}\right\}.  \label{3}
\ee
This expression may be reduced to Eq. (\ref{1}) provided that
\be
|\sigma| \gg \frac{t^2}{\hbar^2v_F^2} . \label{4}
\ee
We remark that the expressions (\ref{2}) and (\ref{3}) are derived for the specific mutual arrangement of the graphene sheets in the bilayer (Bernal stacking). For other geometries, low-energy energy-momentum relations for graphene bilayers take forms quite different from those given by Eq. (\ref{2}), and the approximation for the kinetic term according changes. Nevertheless, the main conclusion concerning the expression for the kinetic term remains justified. This term may be approximated by the expression (\ref{1}) when the density of doped charge carriers on the considered graphene sheet is sufficiently large to satisfy the condition (\ref{4}). On the contrary, when the density of charge carriers on the sheet is small, its low-energy characteristics are strongly affected by the coupling to adjacent layers in the MLG, and the expression (\ref{1}) ceases to be appropriate.

In further analysis we employ a Thomas-Fermi approach to study electric charge and potential distribution over MLG sample. This implies that the interlayer coupling is neglected, and the kinetic term is described by the expression (\ref{1}). Previously, Thomas-Fermi models were successfully  applied to describe electrostatic interactions in graphite intercalate compounds \cite{16,17} as well as to analyze the intrinsic screening in graphene multilayers \cite{8}. However, basing on the above consideration, we stress that obtained results adequately describe  only  those parts of the MLG where the doped carriers densities on the graphene sheets are sufficiently high.

Within the Thomas-Fermi approach, the energy of the doped carriers in the MLG stack includes the kinetic term $K ,$ the term $ U_{int} $ describing electrostatic interactions between graphene layers and the term $ U_0 $ which originates from the interactions between the MLG and the charged substrates. These terms have the form:
\be
K = \sum_i K_i = \frac{2\sqrt\pi}{3}\hbar v_F \sum_i |\sigma_i|^{3/2} \equiv \gamma\sum_i |\sigma_i|^{3/2},   \label{5}
      \ee
\be
U_{int} = - \frac{e^2}{4\epsilon_0} \sum_{i,j} |i - j| \sigma_i\sigma_j ,  \label{6} 
            \ee
\be 
U_0 = -\frac{ e^2}{2\epsilon_0} d \sum_i i\sigma_i  \label{7}                         \ee            
where $\epsilon_0 $ is permittivity of the free space. When the number of layers included in the stack is sufficiently large $(d \ll D) $ the differences between areal densities of doped charge carriers at the adjacent layers are rather small. In this case, we may employ the theory in a continuous limit treating the charge carriers density in the film as a continuous function $ \sigma (z) $ and turning from summation over layers to integration over the interval $ - D/2 \leq z \leq D/2 : $ 
\be
\sum_i \to \int_{-D/2}^{D/2} \frac{dz}{d} \nn .
\ee
For instance, the condition for the electroneutrality of the system within the continuous limit takes on the form:
\be
\sigma_{1s} + \sigma_{2s} +
\int_{-D/2}^{D/2} \frac{dz}{d} \sigma(z) = 0 .  \label{8}
\ee
and the expressions for relevant energies (\ref{5})-(\ref{7}) may be similarly transformed.

As it was demonstrated in the previous work \cite{8}, the expression for the charge-carriers density $ \sigma (z) $ could be derived by minimizing the grand thermodynamic potential for the MLG. Introducing the chemical potential of the system $ \mu $ and combining the expressions for the kinetic and potential energy given by Eqs. (\ref{5})-(\ref{7}), we can write out the following expression for the grand potential $ \Omega: $ 
\begin{align}
\Omega =& \int_{-D/2}^{D/2} \frac{dz}{d} \bigg\{\gamma |\sigma(z)|^{3/2} - \frac{ e^2}{2\epsilon_0}(\sigma_{1s} - \sigma_{2s}) z \sigma(z)  
  \nn\\ & -
  \mu |\sigma(z)|   - \frac{ e^2}{4\epsilon_0}\int_{-D/2}^{D/2}\frac{dz'}{d} \sigma(z) \sigma(z') |z - z'| \bigg\}. \label{9}
\end{align}
The function $ f(z) \equiv |\sigma(z)|^{1/2} $ which minimizes the potential, $ \Omega $ obeys the equation:
\begin{align}
& f(z) - \tilde \mu - \tilde\beta\mbox{sign}[\sigma(z)]
\nn\\ & \times
 \left\{(\sigma_{1s} - \sigma_{2s})  z + \int_{-D/2}^{D/2} \frac{dz'}{d} \sigma(z') |z - z'| \right\} = 0.  \label{10}
\end{align}
Here, $\mbox{sign}(x) $ is the sign function, $ \tilde \mu = 2\mu/3\gamma $ and $\tilde\beta =  e^2/3\gamma\epsilon_0. $ The dimensionless parameter $ \tilde\beta $ measures the ratio of the Coulomb interactions strength to the kinetic energy of the charge carriers in the graphene layers. The Eq. (\ref{10}) is nonlinear with respect to $\sigma(z)$, and this reflects the essential nonlinearity of the Thomas-Fermi theory as applied to MLG systems. The corresponding solution for a conventional material (either conductor or insulator) should be linear with respect to the charge carriers density. The current nonlinearity occurs due to the particular form of the  charge carriers spectra in graphene, which manifests itself in the unusual expression for the kinetic energy given by the Eq. (\ref{1}).

Carrying out successive differentiation  with respect to the variable  $''z''$ and using  the electroneutrality condition given by the Eq. (\ref{8}) one may transform the integral equation (\ref{10}) to the nonlinear differential equation of the second order for the function $ f(z): $
\be
\frac{d^2 f}{dz^2} = \frac{2\tilde\beta}{d} f^2(z)  \label{11}
\ee
with the boundary conditions:
\begin{align}
\frac{df}{dz}\Big|_{z=-D/2} = & -2 \tilde\beta|\sigma_{1s}| ,  \nn\\
\frac{df}{dz}\Big|_{z=D/2} = & 2\tilde\beta|\sigma_{2s}|.  \label{12}
\end{align}
One may note that the function $f(z) $ has a negative slope at $ z= -D/2$ and positive slope at $ z = D/2. $  This gives grounds to conclude that the function $ f(z) $ reaches its minimum at some point $ z = z_0 $ inside the film. Also, one may expect the charge-carriers density $ \sigma(z) $ at the interfaces to take on values whose signs are opposite to those corresponding to the areal charge densities on the substrates $ \sigma_{1s} $ and $ \sigma_{2s}.$
  For certainty, in further analysis we assume that the signs of the areal charge densities on the substrates differ. Then the function $ \sigma (z) $ monotonously increases/decreases over the interval $ -D/2 \leq z \leq D/2 $ depending on the sign of $ \sigma_{1s}, $ and it becomes zero at $ z = z_0 .$ So, this point indicates the location of the layer where no doped carriers are added. Further $z_0 $ is referred to as an electroneutrality level position. It belongs to the real coordinate space and has nothing in common with the neutrality point  on the energy scale which indicates the location of the  Fermi energy of an undoped graphene sheet.  The left and right derivatives of the function $ f(z) $ satisfy the following relation:
\be
\frac{df}{dz}\Big|_{z=z_0 -0} = - \frac{df}{dz}\Big|_{z = z_0 +0} . \label{13}
\ee 
 Again, we remark that the employed Thomas-Fermi model is not justified for the layers with low density of charge carriers. Therefore, the value of $ z_0 $ found within this model may not agree with actual electroneutrality level position. The relation between the two is discussed below.

One cannot analytically solve the differential equation (\ref{11}) following a straightforward way. However, it could be shown (see Appendix) that this equation is equivalent to a conservation law of the form:
\be
\frac{d}{dz}\left\{\frac{1}{2}\left(\frac{df}{dz}\right)^2 - \frac{2\tilde\beta}{3d}f^3(z)\right\} =0.   \label{14}
\ee
Using this conservation law we obtain:
\be 
\left(\frac{df}{dz}\right)^2 = \frac{4\tilde\beta}{3d} f^3(z) + C  \label{15}
\ee
where the constant $C $ is determined by the boundary conditions. We start to analyze the solutions of the Eq. (\ref{15}) by splitting the original range $ -D/2\leq z \leq D/2 $ in parts determined by inequalities  $ -D/2 \leq z < z_0 $ and $ z_0 < z \leq D/2 $ and separately solving this equation over these parts. Employing the boundary conditions (\ref{12}) and introducing a dimensionless parameter $ R_1 $ defined by the expression: 
\be
1 + R_1^3 = \frac{3d\tilde\beta\sigma_{1s}^2}{f^3(-D/2)}  \label{16}
\ee
we may present the solution of the Eq. (\ref{15}) in the form:
\be
\frac{1 + 2z/D}{1 + 2z_0/D} = \frac{\ds\int_r^1 \frac{du}{\sqrt{u^3 + R_1^3}} }{\ds\int_0^1 \frac{du}{\sqrt{u^3 + R_1^3}}} ; \quad  -\frac{D}{2}\leq z <z_0, \label{17}
\ee 
\be
\frac{1 - 2z/D}{1 - 2z_0/D} =\frac{\ds\int_r^a \frac{du}{\sqrt{u^3 + R_1^3}}}{\ds \int_0^a \frac{du}{\sqrt{u^3 + R_1^3}}};  \quad z_0\leq z < \frac{D}{2}. \label{18}
\ee 
In these expressions, $r(2z/D) = f(z)/f(-D/2) $ and $ a = f(D/2)/f(-D/2). $ Also, we may introduce the parameter $R_2 $ defined by the relation similar to that given by the Eq. (\ref{16}):
\be
1 + R_2^3 = \frac{3d\tilde\beta\sigma_{2s}^2}{f^3(D/2)}  \label{19}
\ee
and present the solution of Eq. (\ref{15}) as follows:
\be
\frac{1 + 2z/D}{1 + 2z_0/D} = \frac{\ds\int_{r'}^{1/a} \frac{du}{\sqrt{u^3 + R_2^3}} }{\ds\int_0^{1/a} \frac{du}{\sqrt{u^3 + R_2^3}}} ; \quad  -\frac{D}{2}\leq z <z_0, \label{20}
\ee 
\be
\frac{1 - 2z/D}{1 - 2z_0/D} =\frac{\ds\int_{r'}^1 \frac{du}{\sqrt{u^3 + R_2^3}}}{\ds \int_0^1 \frac{du}{\sqrt{u^3 + R_2^3}}};  \quad z_0\leq z < \frac{D}{2}. \label{21}
\ee 
where $ r' = f(z)\big/ f(D/2). $ The parameters $ R_1 $ and $ R_2 $ are related to each other. Their relation is determined by the Eq. (\ref{13}), and it has the form: $ R_1 = aR_2. $ 
It is  natural to expect that $ a = 1 $ when the charge densities on the substrates are equal in magnitude. In this particular case the electroneutrality level is situated in the middle of MLG film $ (z_0 = 0).$

 Now, it is necessary to clarify the physical meaning of the parameters $ R_{1,2} $ by relating them to certain characteristics describing properties of the considered system. Using Eqs. (\ref{16})-(\ref{21})  one may derive the following expressions:
\be
\left(1 + R_1^3\right)^{1/6} \left\{\int_0^1 \frac{du}{\sqrt{u^3 + R_1^3}} + \int_0^a \frac{du}{\sqrt{u^3 + R_1^3}} \right\} = 2 \Gamma_1,   \label{22}
\ee
\be
\left(1 + R_2^3\right)^{1/6} \left\{\int_0^1 \frac{du}{\sqrt{u^3 + R_2^3}} + \int_0^{1/a} \frac{du}{\sqrt{u^3 + R_2^3}} \right\} = 2 \Gamma_2.   \label{23}
\ee
Here,
\begin{align}
 \Gamma_1 = \left(\frac{\tilde\beta^2 |\sigma_{1s}| D^3}{3d}\right)^{1/3}; %\qquad
\nn\\
\Gamma_2 = \left(\frac{\tilde\beta^2 |\sigma_{2s}| D^3}{3d}\right)^{1/3}. \label{24}
\end{align}

The newly introduced dimensionless parameters $ \Gamma_{1,2} $ are determined by the thickness of the MLG film, the number of the graphene layers included there $ (N = D/d) $ and by the charge carriers densities on the substrates. Also and most importantly, they depend on the parameter $\tilde \beta $ which characterizes the Coulomb interactions between the graphene layers and the substrates as well as interactions between different graphene sheets. All these characteristics are combined into  control parameters, which determine the nature of the MLG screening. When the electrostatic energy predominates $ (\tilde\beta > 1),$   and the areal charge densities at  the interfaces are sufficiently large, the control parameters $ \Gamma_{1,2} $ may take on values significantly greater than $1. $ On the contrary, when the kinetic energy predominates $ (\tilde\beta \ll 1) $ the control parameters $ \Gamma_{1,2} $ should become much smaller than 1 at realistic values of the charge carriers densities of the substrates. As follows from the Eqs. (\ref{22}), (\ref{23}) within the extremely strong electrostatic interactions limit $ (\Gamma_{1,2} \gg 1), $ the parameters $ R_{1,2} $  take on values close to zero. 
In contrast, weak electrostatic interactions $(\Gamma_{1,2} \ll 1) $ result in large values accepted by $R_{1,2}. $ The relation between these  parameters is illustrated in the Fig. 1. One may observe that the crossover between the weak $(\Gamma \ll 1) $ and strong $(\Gamma \gg 1) $  interaction regimes occurs at $ R_{1,2} \sim 1. $ 

\begin{figure}[t] %%% fig. 1
\begin{center}
\includegraphics[width=8.4cm,height=4.3cm]{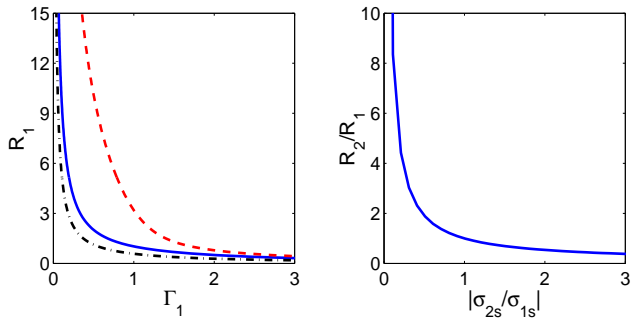}
\caption{(Color online) Left panel: The parameter $ R_1 $ as function of $ \Gamma_1. $ The curves are plotted using Eq. (\ref{22}) and assuming $ a = 10 $ (dashed line), $ a = 1 $ (solid line) and $ a = 0.1 $ (dash-dotted line). For weak electrostatic interactions on the left interface $(\Gamma_1 \ll 1)\ R_1 $ takes on values much greater then 1 being proportional to $ \Gamma_1^{-2}. $ When these interactions are strong $ (\Gamma_1 \gg 1)\ R_1 $ asymptotically approaches zero and could be approximated by $ 1/\Gamma_1. $ Right panel: The ratio $ R_2/R_1 $ versus the ratio of areal charge carriers densities on the substrates. The values taken by $ R_2/R_1 $ vary over a broad range indicating that strong electrostatic interactions between the MLG and the substrate at one interface may  coexist with weak interactions at another interface. When $|\sigma_{1s}| = |\sigma_{2s}|,\ R_1 = R_2 $ regardless of the interaction strength.    
}
 \label{rateI}
\end{center}\end{figure}

Within a strong interactions limit we may approximate:
\be%%gin{align} &
f \left(-\frac{D}{2}\right) = \left(3d\tilde\beta\sigma_{1s}^2\right)^{1/3};  \qquad
%%\nn\\ &
f \left(\frac{D}{2}\right) = \left(3d\tilde\beta\sigma_{2s}^2\right)^{1/3}. \label{25}
\ee%%nd{align}

When electrostatic interactions in the system are weak $(\Gamma_{1,2} \ll 1) \ R_{1,2} \approx \Gamma_{1,2}^{-1}.$ Substituting this approximations into the expressions (\ref{16}), (\ref{19}) we find: 
\be%%gin{align} &
f\left(-\frac{D}{2}\right) = \frac{1}{2} \tilde\beta|\sigma_{1s}|D; \qquad%% \nn\\  &
f\left(\frac{D}{2}\right) = \frac{1}{2} \tilde\beta|\sigma_{2s}|D   .  \label{26}
\ee%%nd{align}
Comparing these asymptotic expressions, we see that the doped carriers densities induced at the MLG interfaces by fixed areal charge densities on the substrates $ \sigma_{1s} $ and $ \sigma_{2s} $ are significantly greater when the Coulomb interactions in the considered system are strong enough for the inequality $ \Gamma_{1,2} > 1 $ to be satisfied. However, even strong Coulomb interactions between the MLG layers may be combined with the weak interactions between the film and the substrates if the areal charge densities on the substrates are small. Thus one or both control parameters $ \Gamma_{1,2} $ may take on small values even provided that $ \tilde\beta > 1. $ 

\begin{figure}[t] %%% fig. 2
\begin{center}
\includegraphics[width=9cm,height=6cm,angle=-90]{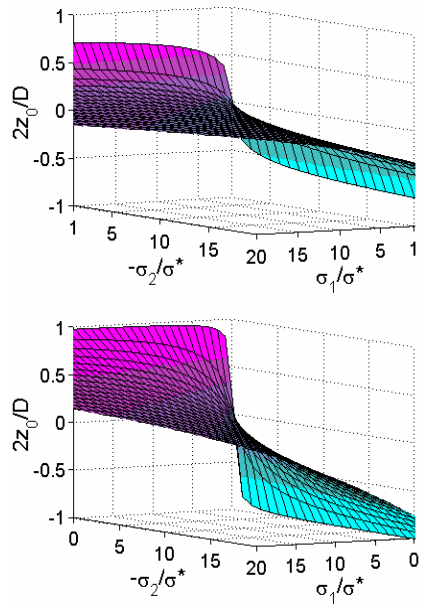}
\caption{(Color online) Location of the electroneutrality level in a twisted MLG film placed in between two charged substrates with the areal charge densities $ \sigma_{1s} > 0 $ and $ \sigma_{2s} < 0 $ at the interfaces $ z = - D/2 $ and $ z = D/2, $ respectively. The presented surfaces are built assuming $ d = 10d = 40\AA, \ \tilde\beta = 10,\ \sigma^* = 10^{-4} \AA^{-2}, $  $ \Gamma_{1,2} > 1 $ (top panel); $\sigma^* =  10^{-2}\AA^{-2},\ \Gamma_{1,2} < 1 $ (lower panel).  
}
 \label{rateI}
\end{center}\end{figure}

Within the accepted approach, the electroneutrality level  location is determined by the relation:
\begin{align}
\frac{4z_0}{D} & = \frac{(1 + R_1^3)^{1/6}}{\Gamma_1}\int_a^1 \frac{du}{\sqrt{u^3 + R_1^3}}
\nn\\ & \equiv
 \frac{(1 + R_2^3)^{1/6}}{\Gamma_2}\int_1^{1/a} \frac{du}{\sqrt{u^3 + R_2^3}}  .
 \label{27}
\end{align}
As follows from this equation, the neutrality level moves through the film as ratio $ a = f(D/2)\big /f(-D/2) $ varies. When $ a< 1 $ it occupies a position shifted to the right from the film center, and in the opposite case $ (a > 1) $ the neutrality level is shifted to  the left. Keeping in mind that the values of the function $ f(z) $ at the interfaces are closely related to the areal charge densities on the substrates in such a way that the ratio $  f(D/2)\big/ f(-D/2) $ accepts values greater/less than one when $ |\sigma_{1s}| $ is less/greater than $|\sigma_{2s}|, $ one may conclude that, in general, the electroneutrality level is situated closer to that interface which is characterized by lower areal charge density on the substrate. The results of numerical solution of the Eq. (\ref{27}) are presented in the Fig. 2, and they confirm the previous qualitative analysis concerning the location of the electroneutrality level inside the MLG film. We remark that at fixed values of $\sigma_{1s}$  and $ \sigma_{2s}, \ z_0 $ keeps closer to the center of the film when the Coulomb interactions are strong $(\Gamma_{1,2} \gg 1) $ than in the opposite case $(\Gamma_{1,2} < 1). $ In the latter case, the electroneutrality level could be moved through the film by varying the ratio $ |\sigma_{1s}|/|\sigma_{2s}| $ between 0.1 and 10. 

\section{ III. Charge distribution}

The doped charge carriers density in the MLG is simply related to the function $ r(z/d).$ Assuming for certainty that $ \sigma_{1s} > 0 $ we obtain:
\be 
\tilde \sigma\left(\frac{2z}{D}\right) = \left\{  \ba{l}\ds -r^2 \left(\frac{2z}{D}\right); \qquad -\frac{D}{2} <z \leq z_0,
  \\ \ds \ \ 
r^2 \left(\frac{2z}{D}\right); \  \qquad  z_0 \leq z \leq \frac{D}{2} .
\ea   \right.  \label{28}
\ee
where $ \tilde \sigma (2z/D) = \sigma (2z/D)\big /f^2(-D/2) $ and the function $ r (2z/D) $ is
 determined  by the Eqs. (\ref{17}),(\ref{18}). These functions are plotted in the Figs. 3 and 4. 
While plotting these curves we assumed for certainty that $ |\sigma_{1s}| $ exceeds $|\sigma_{2s}|,$ so, $ R_2 $ is greater than $ R_1. $ The curves presented in the Fig. 3 are plotted for $ R_1 = 20 $ which corresponds to the weak Coulomb interactions. Under these conditions, $r(2z/D) $ is almost independent of $ R_{1,2}, $ and it may be approximated as follows:
\be
r\left(\frac{2z}{D}\right) = \left\{\ba{l} \ds \frac{2(z_0 - z)/D}{1 + 2z_0/D}; \quad    - \frac{D}{2} \leq z < z_0, 
\\ \\ \ds 
\frac{2a(z - z_0)/D}{1 - 2 z_0/D};  \quad z_0 \leq z \leq     \frac{D}{2}.
\ea \right.  \label{29}
\ee 
which results in the square-law dependence of $ \tilde \sigma $ on $ z/D. $ 
The magnitude of $ \sigma(z) $ rather slowly varies as one moves from the interfaces into the MLG film interior. This indicates that the screening length  of the external charge on the substrates is long  which can be expected since graphene sheets are semimetallic.

\begin{figure}[t] %%% fig. 3
\begin{center}
\includegraphics[width=9cm,height=4.6cm]{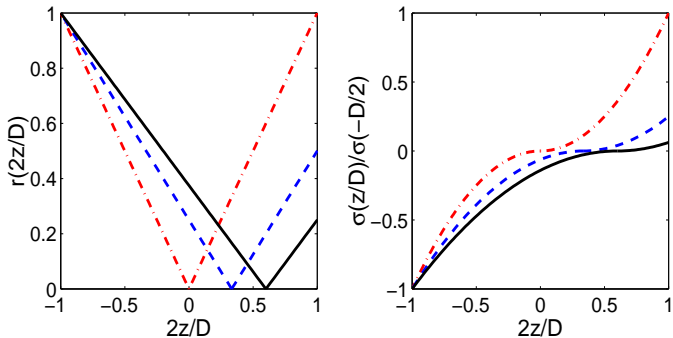}
\caption{(Color online) Ratio $ f(z)\big/f(-D/2) $ (left panel) and $\sigma(z)/\sigma(-D/2) $ (right panel) as a function of the normalized distance into the film. The curves are plotted in accordance with Eqs. (\ref{17}), (\ref{18}) and (\ref{27}) assuming $R_1 = 20 $ and $ a = 1\ (R_2 = 20) $  (dash-dotted lines), $ a = 0.5 \ (R_2 = 40) $ (dashed lines) and $ a = 0.25\ (R_2 = 80) $ (solid lines), respectively.
}
 \label{rateI}
\end{center}\end{figure}

On the contrary, within the strong interaction regime, the doped carriers density magnitude exhibits a steep decrease as we move into the film. This feature is apparent in the dash-dotted curve presented in the right panel of the Fig. 4.
In this case the major portion of the induced charge is concentrated rather close to the interfaces whereas the MLG interior remains nearly neutral. So,  the MLG behaves as a conductor where the external charge is efficiently screened by a surface charge distribution.  Such metallic-like behavior may occur when a sufficiently large amount of charge is put into the substrates  on condition that the graphene layers in the MLG are densely packed, so the electrostatic potential energy predominates over the kinetic term $(\tilde\beta > 1). $

\begin{figure}[t] %%% fig. 4
\begin{center}
\includegraphics[width=9cm,height=4.6cm]{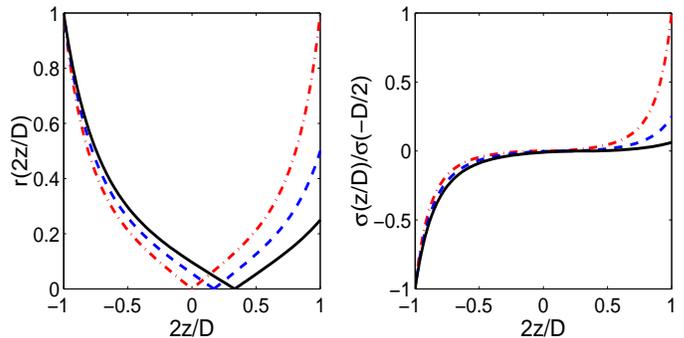}
\caption{(Color online)  Ratio $ f(z)\big /f(-D/2) $ (left panel) and $\sigma(z)/\sigma(-D/2) $ (right panel) as a function of the normalized distance into the film. The curves are plotted in accordance with Eqs. (\ref{17}), (\ref{18}) and (\ref{27}) assuming $R_1 = 0.2 $ and $ a = 1\ (R_2 = 0.2) $ (dash-dotted lines), $ a = 0.5\ (R_2 = 0.4) $ (dashed lines) and $ a = 0.25\ (R_2 = 0.8) $ (solid lines). %% which corresponds to the weak, inter mediate and strong coupling regimes , respectively.
}
 \label{rateI}
\end{center}\end{figure}

The remaining curves in the Fig. 4 are asymmetric. This asymmetry appears due to the difference in the areal charge densities on the substrates. As $|\sigma_{2s}| $ decreases, the Coulomb interactions between the MLG and the substrate at $ z = D/2 $ weakens, thus affecting the distribution of the electric charge in the film near the corresponding interface. It may happen that a strong electrostatic interaction between the film and the substrate on one side of the system is combined with the weak interaction on another side if $|\sigma_{2s}|$ is sufficiently small. Such situation is shown in the Fig. 4 (see the solid line in the right panel). Then the screening lengths at the interfaces significantly differ, and the metallic-like charge  distribution characterizing the portion of the film adjoining the left interface $(z = -D/2) $ fails to appear near the right interface. One may observe a correlation between the position of the electroneutrality level and the relative strength of the MLG interactions with the substrates.  The neutrality level is shifted towards that side of the film where the interaction to the corresponding substrate is weaker, and it approaches to the corresponding interface as the electrostatic interactions further weaken.

\section{IV. Electrostatic potential distribution}

The obtained results enable us to  analyze the electrostatic potential profile across the MLG film. This occurs because the renormalized electrostatic potential $ \tilde\Phi (z) = 2\Phi(z)/3\gamma $  given by the expression:
\begin{align}
e\tilde\Phi(z) = &-\tilde\beta z (\sigma_{1s} - \sigma_{2s})  \mbox{sign} [\sigma(z)] - \tilde\beta\mbox{sign}[\sigma(z)] 
\nn\\ & \times
\int_{-D/2}^{D/2} \frac{dz'}{d}\sigma(z') dz' \label{30}
\end{align}
is simply related  to the function $ f(z). $ Comparing this expression and the Eq. (\ref{10}) we get:
\be
e\tilde\Phi (z) = \tilde\mu - f(z).  \label{31}
\ee
Assuming for certainty that $ \Phi(-D/2) = 0, $ we easily find the corresponding value of the chemical potential $\tilde\mu. $ Substituting the result into  Eq. (\ref{31}) we obtain:
\be
e\tilde\Phi(z) = f\left(-\frac{D}{2}\right) \left[1 + \mbox{sign}(z - z_0)r \left(\frac{2z}{D}\right)\right]. \label{32}
\ee
The profile of the electrostatic potential $\tilde\Phi $ strongly depends on the electrostatic interactions strengths. When the electrostatic interactions in the system are weak $(\Gamma_{1,2} \ll 1),$ we may employ the approximations (\ref{29}) for the function $r(2z/D).$  In this case the electrostatic potential has a linear profile and the potential difference across the film equals: 
\be
\Phi\left(\frac{D}{2}\right) - \Phi\left(-\frac{D}{2}\right) \equiv \Delta\Phi \approx \frac{e D}{4\epsilon_0} (|\sigma_{1s}| + |\sigma_{2s}|).  \label{33}
\ee
So, when the Coulomb interactions are weak and $ \sigma_{1s} = - \sigma_{2s} ,$ the considered system behaves as a parallel-plate capacitor, and the MLG takes on the part of a dielectric material filling the space between the plates and characterized by the dielectric constant $ \kappa = 2 . $

However, when the parameters $\Gamma_{1,2} $ increase due to stronger Coulomb interactions between the film and the substrates the approximation given by the Eq. (\ref{33}) ceases to be valid. For an arbitrary  interactions strength, the electrostatic potential difference 
   exhibits a nonlinear dependence of $ \sigma_{1s} $ and $ \sigma_{2s}. $  Within the strong interactions limit $(R_{1,2} \ll 1) ,$ the potential difference between the interfaces $ \Delta\Phi $ is proportional to $ (|\sigma_{1s}|^{2/3} + |\sigma_{2s}|^{2/3}).$ Assuming that $ |\sigma_{1s}| = |\sigma_{2s}| $ one may consider the system as a capacitor whose differential capacitance varies as $ \Delta\Phi $  changes being proportional to $ (\Delta \Phi)^{1/2}. $
 The  electrostatic potential profiles are presented in the   Fig. 5. When the Coulomb interactions are weak, the potential increases nearly linearly as we move into the MLG film. The stronger are the interactions in the system the more pronounced is the potential change in the vicinities of the interfaces. One may expect that in the limit of the very strong interactions $(R_{1,2} \ll 1)$ almost the whole potential drop should occur near the interfaces leaving the potential nearly constant in the main body of the film.  When $|\sigma_{1s}| \neq |\sigma_{2s}| $ the potential profiles exhibit noticeable asymmetry with respect to the center of the film. This is illustrated in the right panel of the Fig. 5.

\begin{figure}[t] %%% fig. 5
\begin{center}
\includegraphics[width=9cm,height=4.6cm]{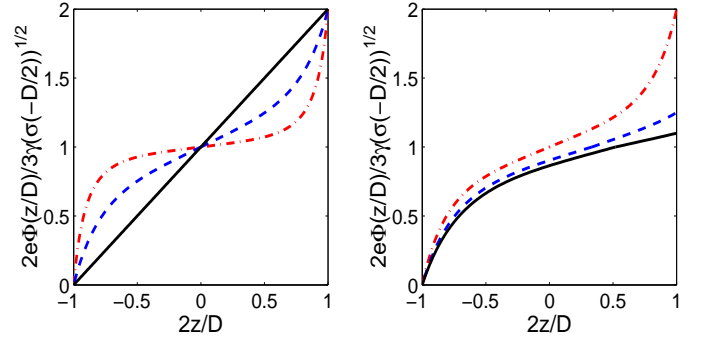}
\caption{(Color online) Spatial profiles of the scaled electrostatic potential inside the MLG film as functions of the normalized distance into the film plotted using Eq. (\ref{32}). Left panel: $ a = 1 $ and $ R_1 = 0.05 $ (dash-dotted line), $R_1 = 0.25 $ (dashed line) and $ R_1 = 20 $ (solid line). Right panel: $R_1 = 0.2 $ and $ a =1 $ (dash-dotted line), $ a = 0.25 $ (dashed line) and $ a = 0.1 $ (solid line).
}
 \label{rateI}
\end{center}\end{figure}

One may  stimulate the switching to the regime characterized by strong Coulomb interactions in the system following two ways. First, one can enhance the areal charge carriers densities on the substrates keeping $ D $ fixed. This would be an appropriate analysis for experiments on a single twisted MLG sample of a certain thickness. The areal charge densities may be varied by varying the voltage applied across the system. Secondly, one may increase the film thickness $ D $ by adding extra graphene sheets to the set. 

To further analyze the electrostatic potential distribution we assume that the areal charge carriers densities on the surfaces of the substrates appear due to the presence of ionized impurities in the substrate materials. For certainty, the impurities  are supposed to be uniformly distributed in the substrates with the volume density $\rho_0 $ which indicates that areal charge densities on the substrates are equal. Then the electrostatic potential inside the substrate layer obeys the Poisson equation which in the considered case is reduced to the form:
\be
\frac{d^2\Phi_s}{dz^2} = - \frac{e}{\epsilon_0}\rho_0.  \label{34}
\ee
  Solving this equation one may find that the total potential difference across the film the total electrostatic potential difference acquires the extra term:
\be
\Delta\Phi_s =  \frac{e\sigma_0^2}{\epsilon_0\rho_0}.  \label{35}%%34}
\ee
where $ \sigma_0 = 2|\sigma_{1s}| = 2|\sigma_{2s}|. $

The  above described specific features of electric charge and potential distribution in MLG samples could be observed if $ \Delta\Phi_s $ is significantly smaller than the contribution to the electrostatic potential difference coming from the graphene film. Within the weak interactions regime, this happens when $ \rho_0 > \sigma_0/D. $
 If the Coulomb interactions are strong the MLG contribution dominates when
\be
\rho_0 > \left(\frac{4}{3d}\tilde\beta^2 \sigma_0^4 \right)^{1/3}.  \label{36}%%35}
\ee
The expression for$ \Delta \Phi_s $  must be changed if we allow for the predominating effect of surface/interface states. This seems  a likely resolution with the culprit being some adsorbed species confined to the interfacial layer.  Then $ \sigma_0 = \rho_0 d_s \ (d_s $ being the characteristic depth for the surface states)  and the contribution from the substrates into the total potential drop accepts the form $ e\sigma_0 d_s/\epsilon_0. $ In this case the predomination of the MLG contribution to the total potential drop becomes easier to reach. When the Coulomb interactions are weak, the graphene contribution  predominates when $ d_s < D $ which is usually the case. When the interactions are  strong, one must require the inequality
\be
d_s <\left(\frac{3d}{4\tilde\beta^2 \sigma_0}\right)^{1/3}   \label{37}%%36}
\ee
to be satisfied to provide the prevalence of the graphene contribution to the potential drop.

\section{V. Discussion}

In the present work we theoretically analyzed characteristic features of electric charge distribution in  MLG stacks placed in between  charged substrates.
  To this purpose, a nonlinear Thomas-Fermi model was developed and employed. Accordingly, the graphene layers in the MLG pack where treated as completely decoupled, and all effects arising due to interlayer hybridization were omitted. As discussed before, this approach is justified only provided that inequality (\ref{4}) is satisfied. Raman scattering experiments on the graphene bilayer within the Bernal geometry have yielded the value of about $0.15 eV $ for the interlayer coupling strength  $ t $ \cite{18}.  Other experiments \cite{19,20,21,22,23}, as well as first-principle computations (see e.g. Ref. \cite{24}), show that this parameter takes on much smaller values when the mutual orientations of adjacent graphene sheets in the pack deviate from certain arrangements corresponding to commensurability between the crystalline lattices of the sheets. Assuming that   $ t \sim 10^{-3} \div 10^{-1} eV $ and estimating the Fermi velocity at a single graphene layer as $ v_F \sim 1.2 \cdot 10^6 m/s $ \cite{24,25}, we may conclude that the condition (\ref{4}) is satisfied for those layers included in the MLG sample which are characterized by charge carriers densities $ \sigma_i $ significantly exceeding  $ \sigma_c\ (\sigma_c = t^2/\hbar^2v_F^2) $ which takes on values of the order of $ 10^{-7} \div 10^{-3} 1/\mathring A^{2}. $ Also, using the presented above estimate for the $ v_F $ we obtain for the dimensionless constant $ \tilde\beta $ the value of the order of $ 10 \ (\tilde\beta = 9.5). $

 Then assuming that $ d$ takes on a value close to the interlayer spacing in graphite $(d \approx 3.347 \mathring A)$ the crossover value of the parameters $ \Gamma_{1,2}\ (\Gamma_{1,2} \sim 1) $ occurs at $| \sigma_{1s}|,|\sigma_{2s}| \approx 3d/2\tilde\beta^2 D^3 \approx 6.5\times 10^{-2}/D^3. $ These areal  densities have dimensions $ \mathring A^{-2} $ if the MLG thickness is expressed in $ \mathring A. $ The minimum value of $ D $ cannot be smaller than $ d, $ so $| \sigma_{1s}|,|\sigma_{2s}| $ need only to exceed $ \sim 2\times 10^{-3} \mathring A^{-2} $ to provide the switching of the system to the metallic-like behavior. We remark that for $ D > d $ these crossover values  may be one or even two orders of magnitude smaller. 
       This gives for the charge density in the MLG near the interfaces between the latter and the substrates the values of the order of $ 10^{-6} \div 10^{-4} 1/\mathring A^{2}. $ For example, assuming that $ D \approx 5d $ and that the carriers densities on both interfaces are equal in magnitude we find $ \sigma_0 = |\sigma(-D/2)| = |\sigma(D/2)| \approx 4\cdot 10^{-6} 1/\mathring A^{2}. $ This significantly exceeds $ \sigma_c $ when the adjacent graphene sheets are twisted away from the angles corresponding commensurability of crystalline lattices so that the coupling parameter has the value of the order of few millielectronvolts. This means that the film regions adjacent to the interfaces may be described using the Thomas-Fermi model. The width of these regions is determined by the ratio $ \sigma_c/\sigma_0. $ It is proportional to $ D(\sigma_c/\sigma_0)^{1/2}$ and could be roughly estimated using Fig. 3 and 4. For $ \sigma_0 \approx 4\cdot 10^{-6} 1/\mathring A^2 $ and $ \sigma_c \approx 10^{-7} 1/\mathring A^2, $ these regions should include about 60\% of the whole film thickness (30\% for each),  and they may be farther expanded by increasing the charge density on the substrates.

So, we may subdivide the MLG sample in three parts. Two of these are regions adjacent to the interfaces where the density of the doped charge carriers is sufficiently high to suppress manifestations of the interlayer coupling. The third (middle) region is characterized by the low density of the charge carriers. One may assert that both actual electroneutality level position and $ z_0 $ determined by Eq. (\ref{27}) are located within this region. By increasing the charge densities on the substrates one may narrow down the width of this region, thus bringing the two closer.

 Certainly, the areal charge carriers densities at the surfaces of substrates   of the order of $ 2 \cdot 10^{-3} 1/\mathring A^2 $
 are within the experimentally accessible range. However, specific features originating from the special properties of graphene could be manifested only provided that the contributions from the substrates to the total  electrostatic potential distribution in the system are small compared to the contribution from the MLG. As follows from the Eqs. (\ref{35}),(\ref{36}) this occurs when the volume density of charge carriers uniformly distributed in the substrates $ \rho_0$ takes on values of the order of $ 10^{20} 1/cm^3 $  which is quite large.  The situation is much better in the case when the areal charge carriers densities at the substrate surfaces appear due to the effect of surface states with the characteristic depths $ d_{1s} $ and $d_{2s}.$  Using Eq. (\ref{37}), we may estimate $d_{1s},d_{2s} \sim 1nm $ at $ \sigma_{1s},\sigma_{2s} \sim 10^{-3} \mathring A^{-2}. $ This is quite reasonable. The substrate only gets to dominate if the characteristic depths become significantly longer than these estimated above. So, the above discussed effects originating from the particular charge carriers spectra in graphene  are likely to be accessible for experimental observations.

 Again, we remark that the the theory developed in the present work used a conical dispersion relation for the charge carriers on the graphene sheets which results in the specific expression for the kinetic energy density. If the Dirac cone is distorted (which may happen in a practical MLG sample) this should bring changes into the expression for the kinetic energy dependence of the carriers density. If these changes are significant, it may significantly modify the electric charge and potential distribution over the film. However, it is likely, that these distortions are more pronounced at low energies, so they bring only  rather small corrections to the kinetic energy density, provided that the local Fermi energies are sufficiently large. Then these corrections could be omitted, and the main equations and results retain their form.

Electronic properties of twisted MLG systems have attracted a significant interest of the research community which has resulted in intensive theoretical studies using several approaches and techniques \cite{13}. It is commonly acknowledged that the hybridization between the layers in these systems (although weak) still may influence their low-energy electronic characteristics and bring changes in some observables. For instance, it was predicted that in such systems Fermi velocities should be reduced by a factor depending on the twist angle. However, recent experiments aiming at observation of these effects give inconsistent results \cite{26,27}. An obvious explanation of the discrepancies is that manifestations of interlayer hybridization in twisted MLG samples  are weak, and it may be extremely difficult to distinguish them against the background created by various external factors. Also, it seems likely that interlayer hybridization in MLG samples could vary as one moves through the system. It may happen that the layers in a certain portion of the sample are significantly stronger coupled to their neighbors than those located in another  part of this pack. While measuring the response of the whole sample, the contributions from the stronger coupled layers could predominate over the contributions from weaker coupled ones, thus leading to discrepancies between experimental results and theoretical predictions. 

When the twisted MLG sample is placed in between two charged substrates, the electric charge density appears in the film. All graphene layers except of those situated near the electroneutrality level bear electric charge. Strong Coulomb interactions between these charged levels may suppress weak effects originating from interlayer hybridization. The situation is different in the vicinity of the neutrality level. The influence of hybridization between the layers located here is not suppressed, and many-body effects may be manifested.   By strengthening the electric field applied  across the MLG sample or by  increasing the carriers densities on the adjoining substrates one may narrow the region where the neutrality point is located  down to a few percent of the film thickness. Then one may move the charge neutrality locus through the film by tuning the charge densities on the substrates or varying the electrostatic potential applied across the sample thus creating opportunities to separate out a few graphene layers and study the effects of their hybridization with the neighboring layers and variations of these effects over the sample. This could provide means for better understanding of electronic properties of twisted multilayered graphenes.

\section{Appendix}

Here, we present the mathematics which allows us to derive the Eq. (\ref{11}). We start from the Eq. (\ref{8}). By multiplying both sides of this equation by $df/dz $ we get:
\be
\frac{df}{dz}\frac{d^2f}{dz^2} = \frac{df}{dz}\frac{2\tilde\beta}{d} f^2(z). \label{38}
\ee
Then we use the relations:
\be
\frac{df}{dz}\frac{d^2f}{dz^2}= \frac{1}{2}\frac{d}{dz}
\left[\left(\frac{df}{dz}\right)^2\right] \label{39}
\ee
and
\be
f^2(z) \frac{df}{dz} = \frac{1}{3}\frac{d}{dz}(f^3(z)).  \label{40}
\ee
Substituting these identities into the Eq. (\ref{36}) we obtain the conservation law (\ref{11}) which is used in the main body of the work.

\section
{Acknowledgments} 
We  thank  G. M. Zimbovsky for help with the 
manuscript. This work was partly supported  by  NSF-DMR-PREM 0353730.

%\begin{widetext}  \end{widetext}

\end{document}